\pdfoutput=1
\documentclass[pre,aps,showpacs,floatfix,nofootinbib,superscriptaddress,10pt]{revtex4-2}
\usepackage{subfigure}
\usepackage{amssymb}
\usepackage{amsfonts}
\usepackage{amsmath}
\usepackage{amsthm}
\usepackage{epsfig}
\usepackage{graphicx}
\usepackage{orcidlink}
\AtBeginDocument{%
  \renewcommand{\url}[1]{}%
  \providecommand{\issn}[1]{}%
  \renewcommand{\issn}[1]{}%
}

\usepackage{xpatch}

\makeatletter
\AtBeginDocument{%
  \def\ISSN#1{}%
  \def\issn#1{}%
}
\makeatother

\usepackage[utf8]{inputenc}

\hypersetup{colorlinks=true,
 	linkcolor=blue,
 	urlcolor=blue,
 	citecolor=blue,
 	pdfhighlight=/N
 }
 \usepackage{comment}
 \usepackage[normalem]{ulem}

\begin{document}

\title{Physics-based logarithmic description of electrostatic field enhancement in hemisphere-on-cylindrical-post structures for high-field applications}

 \author{Thiago A. de Assis}
\email{thiagoaa@id.uff.br (Corresponding author)}
   \affiliation{Instituto de F\'{\i}sica, Universidade Federal Fluminense,
  Avenida Litor\^{a}nea s/n, 24210-340, Niter\'{o}i, Rio de Janeiro, Brazil}

 \author{Fernando F. Dall'Agnol}
 \email{fernando.dallagnol@ufsc.br}
 \affiliation{Department of Exact Sciences and Education, Universidade Federal de Santa Catarina, Campus Blumenau, Rua Marechal Rondon, 880, Salto do Norte, Blumenau 89065-200, Santa Catarina, Brazil}

\author{Richard G. Forbes}
\email{r.forbes@surrey.ac.uk}
\affiliation{Quantum Sciences Group, School of Mathematics and Physics, University of Surrey, Guildford, Surrey GU2 7XH, UK}

\begin{abstract}

Electrostatic (ES) field enhancement at sharp conducting structures plays a central role in lightning protection, corona discharge, electrical breakdown in vacuum (for example in particle accelerators), and more generally in technological applications of field electron emitters. A canonical geometry for studying this ES effect is the hemisphere-on-cylindrical-post (HCP) model, in the regime where the structure stands on a planar conductor of large lateral extent, and a large gap exists between the structure and the counter-electrode. A parameter of major interest is the apex field enhancement factor (apex FEF) $\gamma_{\mathrm{a}}$ [$\equiv$ ``apex ES field"/``background ES field"].  For this (and other) structures, a formula for $\gamma_{\mathrm{a}}$ can be written in the form $\gamma_{\mathrm{a}} = c_{\mathrm{a}}(\sigma_{\mathrm{a}}) \times \sigma_{\mathrm{a}}$, where the apex sharpness ratio $\sigma_{\mathrm{a}}$ is given by the ratio (``post-height"/``apex-radius-of-curvature"), and the apex sharpness coefficient $c_{\mathrm{a}}(\sigma_{\mathrm{a}})$ depends on the post shape. For the HCP model, no exact analytical formulas for $c_{\mathrm{a}}(\sigma_{\mathrm{a}})$ or $\gamma_{\mathrm{a}}$ are currently known. (Quite possibly none exist.) This paper develops a compact analytical approximation for $c_{\mathrm{a}}(\sigma_{\mathrm{a}})$ and hence for $\gamma_{\mathrm{a}}$. This compact formula has a ``logarithmic-style" structure, rather than the ``power-style" structure used in previous approximations. When compared with precise finite-element analyses of the HCP model, over the computationally accessible range $1 \leq \sigma_{\mathrm{a}} \leq 1000$, this ``logarithmic-style" formula has a maximum error-magnitude of $0.15 \ \%$, which is significantly better than older approximations. Details of our methodology of derivation, and some consequences of the resulting formula, are discussed. 

\end{abstract}

\maketitle

\section{Introduction}

\subsection{General background}

It is well known informally that the local electrostatic (ES) field $E_{\rm{a}}$ at the apex of a real-world lightning rod is significantly higher in magnitude than the background ES field $E_{\rm{back}}$ associated with the negative electric charge of Planet Earth. This can be called the \textit{electrostatic lightning-rod effect} and is a particular example of the more general phenomenon of electrostatic field enhancement, whereby the local ES field is relatively high in magnitude over sharp features of electrically charged objects.

This ES effect needs to be distinguished from the electrical effects induced when the electric component of a travelling electromagnetic (EM) wave acts on a rod-like structure. This EM effect is best called the electromagnetic antenna effect and involves physics different from the ES lightning-rod effect. Confusion between the two effects sometimes occurs in the literature.

If the lightning rod or other pointed feature is in air, and if the apex ES field is sufficiently high in magnitude (typically around $3\times 10^{6}\,\mathrm{V/m}$), but somewhat variable, depending on atmospheric conditions, then a corona discharge \cite{Corona} occurs. This is the explanation of some atmospheric phenomena, especially St Elmo's fire, and is considered by some to play a significant role in the physics of lightning protection. If the (negative) apex ES field were to reach a magnitude of around $2\times 10^{9}\,\mathrm{V/m}$, then field electron emission (FE) \cite{FELit} could occur; but the normal assumption is that (in a gas) corona discharge would occur first.

Electrostatic field enhancement of this kind can be quantified by a (dimensionless) \textit{apex field enhancement factor} (apex FEF), denoted here by $\gamma_{\rm{a}}$ and defined as the ratio of the local ES field at the emitter apex, $E_{\rm{a}}$, to the background ES field $E_{\rm{back}}$:
\begin{equation}
\gamma_{\rm{a}} \equiv E_{\rm{a}} / E_{\rm{back}}.
\label{eq:gamma_def}
\end{equation}

The value of $\gamma_{\rm{a}}$ depends strongly on the geometry of the system under discussion. There is both scientific and technological interest in understanding the related physics and in deriving formulas for $\gamma_{\rm{a}}$ for various model geometries.

\subsection{Theoretical background}

There is particular theoretical interest in situations that can be electrostatically modeled as a cylindrically symmetric conducting post, with a rounded apex, standing on one of a pair of parallel planar conducting plates of large lateral extent, in the ``large-gap regime" where the post height is very much less than the plate separation.

For such situations, there is strong evidence (see Appendix 1) that a physically important parameter is the \textit{apex sharpness ratio (ASR)} $\sigma_{\rm{a}}$, defined as the post height $h$ divided by the apex radius of curvature $r_{\rm{a}}$, i.e. by
\begin{equation}
\sigma_{\rm{a}} \equiv h/r_{\rm{a}}.
\label{sigma}
\end{equation}

Note that this parameter $\sigma_{\rm{a}}$ needs to be distinguished from the \textit{aspect ratio}, which -- for cylindrically symmetric models -- is the ratio (``protrusion height''/``base radius''). For the HCP model discussed below, these parameters have the same value, but for all other models of interest the values are different.

There is also suggestive evidence that, for cylindrically symmetric models, formulae for the apex FEF $\gamma_{\rm{a}}$ can often or always be written in the useful form 
\begin{equation}
\gamma_{\rm{a}} = c_{\rm{a}}(\sigma_{\rm{a}})\, \cdot \sigma_{\rm{a}},
\label{eq:gamma_gen}
\end{equation}
where the \textit{apex sharpness coefficient} $c_{\rm{a}}$ is a function of $\sigma_{\rm{a}}$. Characteristically different post shapes (and different mathematical approximations for a given post shape) will yield different expressions for $c_{\rm{a}}(\sigma_{\rm{a}})$ and hence for $\gamma_{\rm{a}}(\sigma_{\rm{a}}).$

\subsection{Specific aims}

\begin{figure*}[t]
\centering
\includegraphics[width=0.7\linewidth]{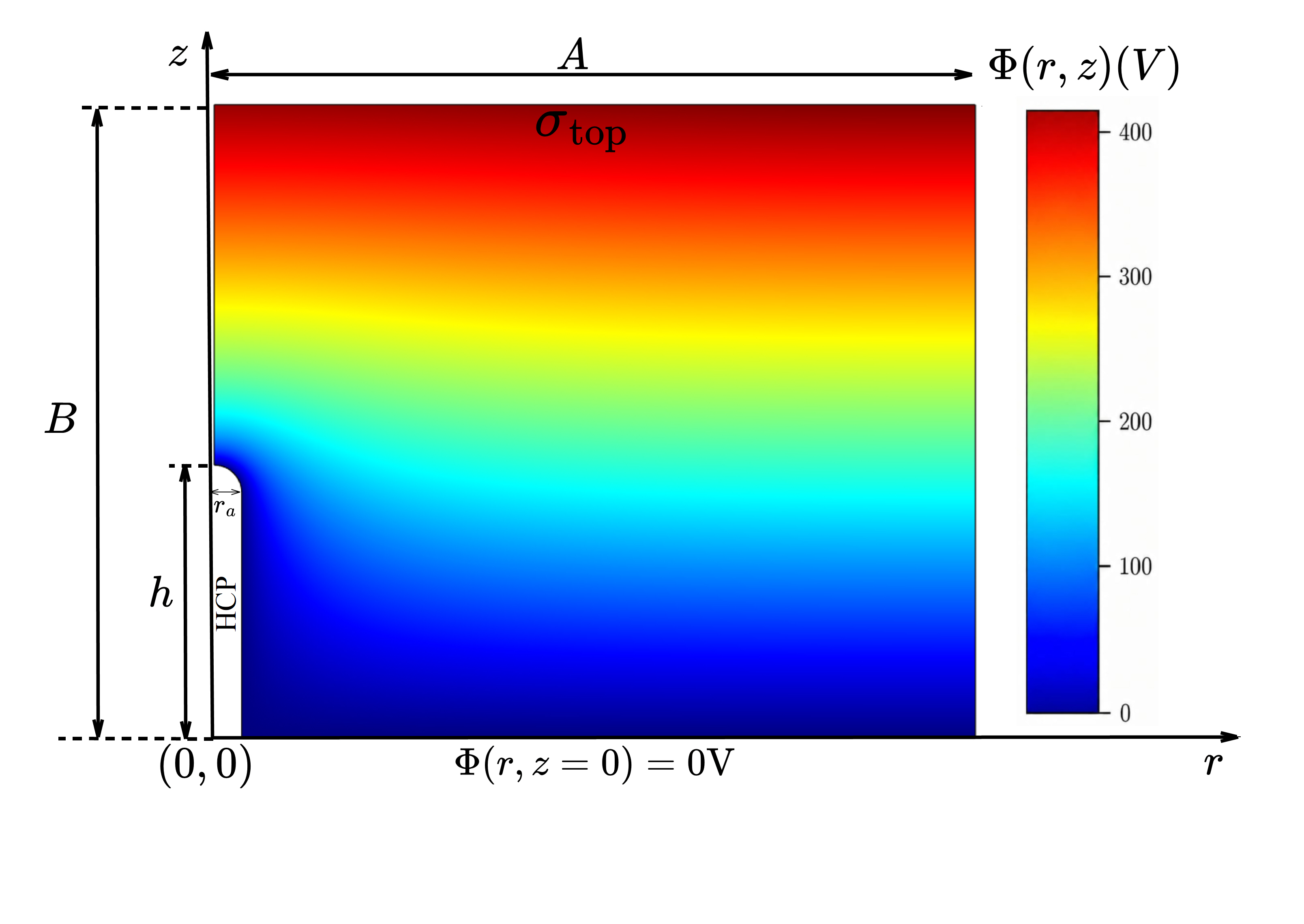}
\vspace{-1.0cm}
\caption{
Axisymmetric finite-element simulation domain used to compute the
electrostatic potential $\Phi(r,z)$ and the apex field enhancement
factor for the hemisphere-on-cylindrical-post (HCP) model in the
large-gap regime. The HCP emitter, of height $h$ and apex radius
$r_{\mathrm a}$, is mounted on a grounded conducting plane. The
computational domain is bounded radially at $r=A$ and vertically at
$z=B$, with $B \gg h$. A Neumann boundary condition corresponding to a
uniform surface charge density
$\sigma_{\mathrm{top}}=-\varepsilon_0E_{\mathrm{back}}$ is imposed at
the top boundary, generating the background electrostatic field
$E_{\mathrm{back}}$. The colour map shows the electrostatic potential
distribution obtained from the axisymmetric finite-element solution of
Laplace’s equation.}
\label{fig:HCP_domain}
\end{figure*}

A shape of particular interest is the cylindrical post with a hemispherical cap (the so-called ``HCP model"), in the large-gap regime (see Fig.~\ref{fig:HCP_domain} for a two-dimensional representation). This HCP model is a good ``starting-point" geometrical model for a real-world lightning rod (if the cap can be approximated as hemispherical and if any surface roughness of the apex is ignored). In microscale/nanoscale contexts, it is also a good starting-point model for whisker-like protrusions and for upright carbon nanotubes.

The HCP model thus has applications in the theory of electron-emission devices such as large-area field electron sources \cite{evtukh2015}, and also applications in the theory of electrical breakdown in vacuum in several technological contexts, including electron beam instruments, medical X-ray scanners, and large-scale particle accelerators.

Consequently, especially in these technological contexts, the electrostatics of the HCP model has attracted much theoretical attention over the last 60 years, and the physics, algebra and numerics of this model are now reasonably well understood. A recent review~\cite{ESReview} by the present authors includes summaries of past work on, and of the present state of, the HCP model.

The electrostatics theory developed in the above ``microscopic'' contexts is in fact of general applicability, including to many terrestrial/atmospheric electrical phenomena~\cite{Corona, Lightning}. However, the related knowledge does not usually appear in general Electricity and Magnetism textbooks, and appears not yet to be well known to some communities where it would be relevant.

The primary aims of this paper are twofold. The first is to report a new, more accurate, algebraic expression for the apex sharpness coefficent (and hence the apex-FEF) for the HCP model. This expression has been derived by least-squares fitting to numerical results similar to those the authors reported previously \cite{FEF2021,ESReview}, but now uses a ``logarithmic-style" algebraic form that resembles the algebraic forms used to predict $c_{\rm{a}}(\sigma_{\rm{a}})$ and $\gamma_{\rm{a}}(\sigma_{\rm{a}})$ for the hemi-prolate-spheroidal (HPS) model of a conducting post (e.g., \cite{Kom91, RGFIVNC26HPS}), rather than the ``power-style" algebraic forms previously used for the HCP model.

The second aim is to provide the basis for a later attempt to transfer basic knowledge about electrostatic field enhancement and the HCP model into the lightning protection community. In particular, it appears to the present authors that this community knows that the degree of field enhancement is influenced both by the height ($h$) of the lightning rod, and by the radius of curvature ($r_{\rm{a}}$) of its apex, but does not yet widely appreciate that a key scientific parameter in the theory of ES field enhancement is the apex sharpness ratio $h/r_{\rm{a}}$.

The present authors have a long-standing interest in the physics and electrostatics of the HCP model. However, we note that the work reported here has partially been stimulated by the paper of Gomes and Parus \cite{GomesParus}. In their discussion relating to the lightning-rod effect, in the context of lightning protection, these authors have apparently used a field enhancement formula derived by Biswas~\cite{BiswasFEF} -- although Biswas does not in fact claim that his formula applies  exactly to the HCP model. We regard the Gomes and Parus use of the Biswas formula as an approximation that is fair but may not be exactly applicable.  This has stimulated us to look again at the formula we previously derived for the HCP model.

Derived classical field-enhancement formulae are valid for both positive and negative fields. However, this paper specifically considers the negative-field situation, on the grounds that this better suits the technological contexts under discussion.


\subsection{Some earlier field-enhancement formulae}

As far as is known, the only protrusion shapes for which exact analytical formulae are known are the various forms of the hemi-ellipsoid on a plane. In practice, in field-enhancement contexts, only the formulae for the cylindrically symmetric ``hemi-prolate-spheroid on a plane (HPS) model" are normally discussed. This HPS model includes the special case of a hemisphere on a plane, for which there is a long-established and easily provable exact result that $c_{\rm{a}}(1) = 3$ precisely (see Appendix 2).

The HPS model (formerly called the ``hemi-ellipsoid on a plane (HEP) model") and its background are discussed in our review~\cite{ESReview} and have been discussed further in a recent conference presentation \cite{RGFIVNC26HPS}. There is an exact solution, but for $\sigma_{\rm{a}}$ values that are sufficiently large (greater than about 40) this reduces to the simpler formula (see eq. (42) in \cite{ESReview})
\begin{equation}
c_{\rm{a}}(\sigma_{\rm{a}})(\rm{HPS}) \approx 1/ \left[ \tfrac{1}{2}\ln(4 \sigma_{\rm{a}}) - 1 \right].
\label{eq:hep_asym}
\end{equation}

By contrast, formulae for the apex FEF for the HCP model have historically not involved logarithmic expressions. From the 1960s onwards, many different formulas have been generated, using a variety of different theoretical models and methods. These include: (a) the ``floating sphere at emitter-plate potential" \cite{Gom57,Gom58,Vibrans1964a,Vibrans1964,Miller1,Lat81,Nic93,Xu95,FilipV01, Forbes2003,FilipV04, Wang04,Pog10,RFJAP2016}; (b) complex electrostatics modeling \cite{Xanthakis}; (c) numerical line-charge-based models \cite{Vibrans1964a,Vibrans1964,Podenok06,BiswasFEF}; (d) numerical finite-element methods
\cite{Edgcombe2001,Edgcombe3,Edgcombe2,BonardPRL2002,ZhuRef,Silva2005,ZENG2009,Roveri16,JVSTB2019,ESReview,FEF2021}; and (e) a numerical boundary-element method \cite{RBowring}. 
Refs \cite{Forbes2003} and \cite{ESReview} provide partial reviews.

Some references attribute enhancement formulae to Rohrbach, but no such formulas appear in the technical report usually cited \cite{Roh71}.

It was shown in an earlier paper by two of us \cite{JVSTB2019} that several of the formulae derived in past work can be put in the general form
\begin{equation}
\gamma_{\rm{a}} \approx a \left[b + \left(\sigma_{\rm{a}}\right)^{c}\right]^d,
\label{eq:old_fit}
\end{equation}
where $a$, $b$, $c$, and $d$ are fitting parameters that were determined by regression over a finite range of values of $\sigma_{\rm{a}}$, usually lying between 1 and 1000. For the earlier papers of most interest here, the resulting parameter values are shown in Table I below.

In this earlier approach, no attempt was made to incorporate the known lower bound that $c_{\rm{a}}(1) = 3$.

\section{High-Precision Numerical Reference Data for the HCP Model}
\label{sec:reference_data}
\noindent

Our review~\cite{ESReview} describes careful procedures used to establish the accuracy of finite-element-method calculations made using the COMSOL package. A numerical limitation is that demands on memory and calculation time become unmanageably large as $\sigma_{\rm{a}}$ gets larger beyond a certain limit. Thus, we currently confine our simulations to the range $1 \le \sigma_{\rm{a}} \le 1000$, which should be good enough for most practical purposes.

\begin{figure*}[t]
	\centering
	\includegraphics[width=0.7\linewidth]{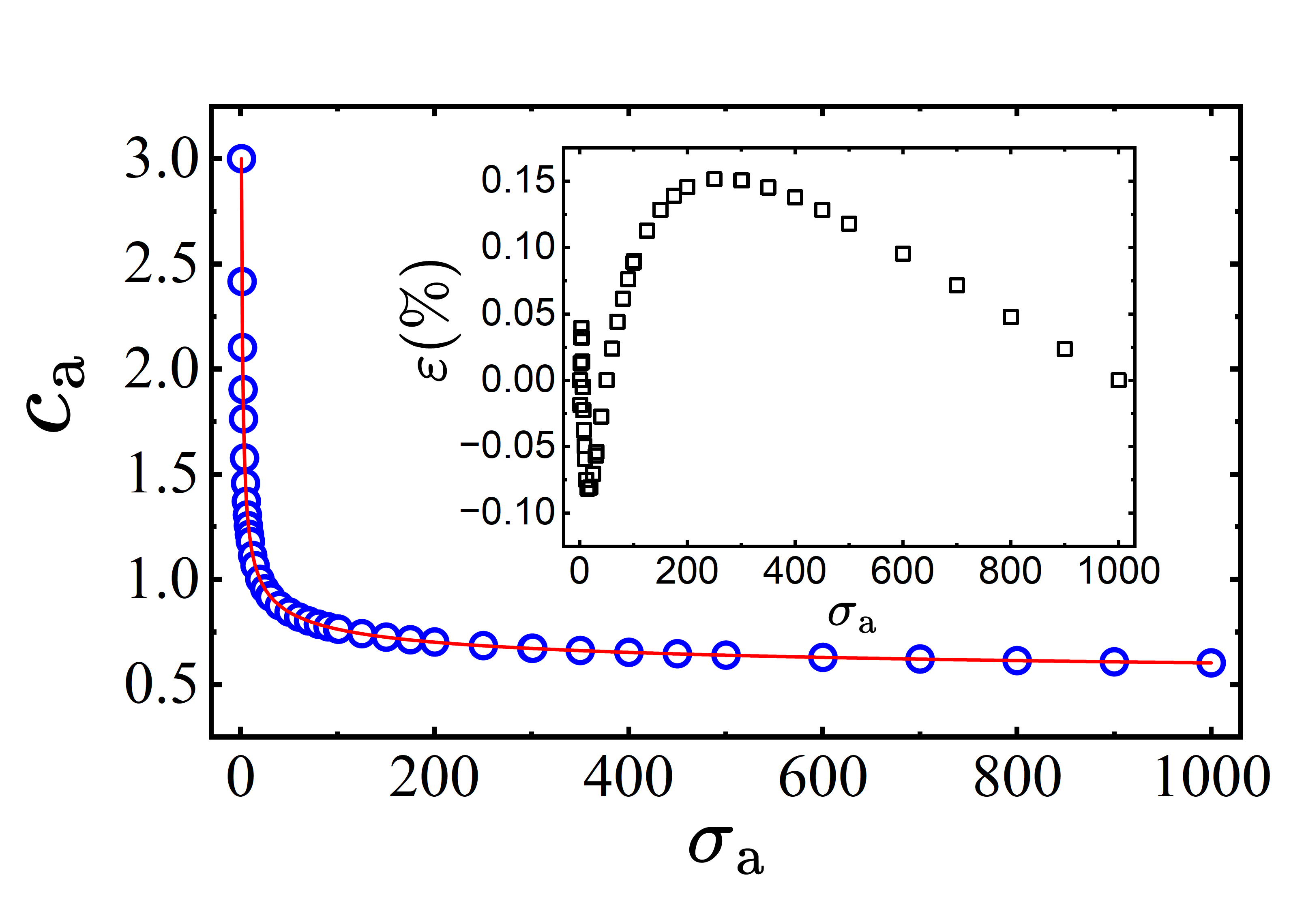}~ ~
    \caption{Geometric coefficient $c_{\mathrm{a}}(\sigma_{\mathrm{a}})=
\gamma_{\mathrm{a}}/\sigma_{\mathrm{a}}$ as a function of the apex sharpness ratio
$\sigma_{\mathrm{a}}=h/r_{\mathrm{a}}$ for the HCP model in the large-gap regime.
Blue circles represent high-precision finite-element results obtained using the MDD technique, with an estimated relative numerical uncertainty in $\gamma_{\mathrm{a}}$ of order $10^{-5} \ \%$. The numerical data show a smooth and monotonic decrease of $c_{\mathrm{a}}$ from the exact hemispherical value $c_{\mathrm{a}}(1)=3$ toward a regime of progressively weaker variation as $\sigma_{\mathrm{a}}$ increases. The solid red curve is the least-squares fit of the analytical expression defined by Eqs.~\eqref{eq:ca_interp} and \eqref{eq:F_def} with fitted parameters as discussed in the text. The inset shows the relative percentage deviation $\varepsilon(\sigma_{\mathrm{a}})$ between the fitted analytical expression and
the numerical data. The deviation remains within approximately $\pm 0.15 \ \%$ over the full interval $1 \leq \sigma_{\mathrm{a}} \leq 10^{3}$, confirming that the proposed formula reproduces both the rapid variation near the hemispherical limit and the logarithmically slow evolution observed for large values of $\sigma_{\mathrm{a}}$.}
\label{fig:ca_sigma}
\end{figure*}

For this $\sigma_{\rm{a}}$ range, methodology has been described \cite{ESReview,FEF2021,MDD2023} that, when predicting $\gamma_{\rm{a}}$,  can achieve accuracy better than a required input percentage level, typically $0.1 \ \%$ or $0.001 \ \%$, but slightly better if needed. For the present and previous work involving fitting-formula comparisons we have used an accuracy of $10^{-5}$ \ \%. Finite-element calculations at this level generate values for $\gamma_{\rm{a}}(\sigma_{\rm{a}})$ and thus the related behaviour of $c_{\rm{a}}(\sigma_{\rm{a}})$, which is illustrated in Fig.~\ref{fig:ca_sigma}.

%

Specifically, the high-precision numerical data for the HCP-model geometry were derived as follows. The electrostatic (ES) potential $\mathit{\Phi}(r,z)$ was computed using an axi-symmetric, finite-element numerical solution of Laplace’s equation in cylindrical coordinates, where $r$ denotes the radial distance from the symmetry axis and $z$ the vertical coordinate measured from the base-plane.

As illustrated in Fig.~\ref{fig:HCP_domain}, the effective simulation domain comprised the volume between (a) the HCP-model emitter mounted on a base-plane and (b) a surrounding cylindrical box, of large but finite height, that had the same symmetry axis as the emitter. The radial boundary was placed at a distance $A$ from the symmetry axis and the planar top boundary was placed at a distance $B$ from the base-plane. The value of $B$ was chosen such that $B \gg h$, corresponding to the large-gap (distant upper-plate) regime. The HCP-model emitter and the base-plane were treated as perfect conductors. For simplicity, the usual assumption was made that all relevant surfaces have the same local work function everywhere, and the electrostatic potential just outside the emitter and the base plane was taken equal to $\mathit{\Phi} = 0$ V.

A Neumann-type boundary condition corresponding to a uniform (positive) surface charge density $\sigma_{\mathrm{top}} = - \varepsilon_0 E_{\mathrm{back}}$ (where $\varepsilon_0$ is the electric constant) was imposed on the top boundary, at $z=B$. Together with the induced (negative) charge on the base-plane, this would generate a negative uniform background electrostatic field $E_{\mathrm{back}}$ in the absence of the protrusion. The lateral boundary was treated as a symmetry feature, by imposing the conditions that the ES field be parallel to the boundary and normal to the base-plane. These requirements create a two-dimmensional geometrical domain in which the three-dimensional Laplace equation is solved by finite-element methods. In accordance with the Minimal Domain Dimension (MMD) criterion~\cite{ESReview,MDD2023}, the values of $A$ and $B$ were selected so that boundary-induced perturbations of the apex field enhancement factor were insignificant. Mesh-refinement and domain-size convergence tests were performed until the relative percentage error in the computed apex-FEF was controlled at the level of $10^{-5} \ \%$.

The blue circles in Fig. 2 illustrate the resulting behaviour of ${c_{\rm{a}} (\sigma_{\rm{a}})} =\gamma_{\rm{a}}/{\sigma_{\rm{a}}}$ over the interval $1 \le \sigma_{\rm{a}} \le 1000$. As expected intuitively, a smooth monotonic decrease in $c_{\rm a}$ with increasing $\sigma_{\rm a}$ was found.

\section{HCP-model Formula Construction}
\label{sec:formula_construction}

\subsection{Introduction}

The numerical results shown in Fig.~\ref{fig:ca_sigma} show that the geometric coefficient $c_{\mathrm{a}}(\sigma_{\mathrm{a}})$ is a smooth and strictly decreasing function of the apex sharpness ratio $\sigma_{\mathrm{a}}$, throughout the interval $1 \leq \sigma_{\mathrm{a}} \leq 10^{3}$. The numerical behavior reveals two regimes. Near the hemispherical limit, $\sigma_{\mathrm{a}}=1$, the coefficient $c_{\mathrm{a}}$ decreases rapidly with $\sigma_{\mathrm{a}}$, indicating that the apex field is highly sensitive to changes in the overall emitter geometry. By contrast, for large values of $\sigma_{\mathrm{a}}$, the variation becomes progressively weaker. In this regime, the numerical data suggest that the dependence of
$c_{\mathrm{a}}$ on $\sigma_{\mathrm{a}}$ is quasi-logarithmic.

This behaviour can be understood physically as follows. For relatively small
values of $\sigma_{\mathrm{a}}$, the electrostatic field near the apex is
influenced by the geometry of the entire protrusion, including the cylindrical
shaft and the conducting plane on which it stands. Consequently, changes in
height produce comparatively large changes in the apex field enhancement.
As $\sigma_{\mathrm{a}}$ increases, the field determination becomes increasingly localized to
near the apex region, and the influence of remote geometric features decreases.
The apex field is then governed primarily by the local geometry at the top of
the post, so that further increases in $\sigma_{\mathrm{a}}$ produces only
incremental changes in the coefficient $c_{\mathrm{a}}$.

The analytical construction developed below is designed to reproduce these
features while incorporating the exact hemispherical limit $c_{\mathrm{a}}(1)=3$.

\subsection{Interpolation structure}

The numerical data suggest that $c_{\mathrm{a}}(\sigma_{\mathrm{a}})$ may be
constructed as a smooth interpolation between exact values at
$\sigma_{\mathrm{a}}=1$ and at a specified large-$\sigma_{\rm{a}}$ value, in our case
$\sigma_{\mathrm{a}} = 1000$.

We therefore use an expression of the form
\begin{equation}
c_{\mathrm{a}}(\sigma_{\mathrm{a}})
=
c_{\mathrm{L}}
+
\frac{3-c_{\mathrm{L}}}
{1+F(\sigma_{\mathrm{a}})},
\label{eq:ca_interp}
\end{equation}
where $F(\sigma_{\mathrm{a}})$ is a non-negative ``transition function" satisfying the conditions
\begin{equation}
F(1)=0,
\end{equation}
\begin{equation}
c_{\mathrm{a}}(1000)
=
c_{\mathrm{L}}
+
\frac{3-c_{\mathrm{L}}}
{1+F(1000)}.
\label{eq:ca_interp2}
\end{equation}

The parameter $c_{\mathrm{L}}$ is a constant that relates to the large-$\sigma_{\mathrm{a}}$
behaviour of the parametrization. Its value is decided by the choice of the function
$F(\sigma_{\mathrm{a}})$ and is determined as described in Appendix 3.
 
Equation~\eqref{eq:ca_interp} automatically yields $c_{\mathrm{a}}(1)=3$, independently of the value of $c_{\mathrm{L}}$.
If $F(\sigma_{\mathrm{a}})$ is chosen as a strictly increasing function,
then $c_{\mathrm{a}}(\sigma_{\mathrm{a}})$ necessarily decfeases monotonically.

\subsection{Choice of the transition function}

The next step is to choose a functional form for $F(\sigma_{\mathrm{a}})$. The numerical data in Fig.~\ref{fig:ca_sigma} show that, in our simulations, the transition from the exact hemispherical value $c_{\mathrm{a}}(1)=3$ to the weakly varying high-$\sigma_{\mathrm{a}}$
regime is distributed over approximately three decades in
$\sigma_{\mathrm{a}}$. This broad transition suggests a dependence that
grows much more slowly than any power of $\sigma_{\mathrm{a}}$.

Additional support for this conclusion is provided by the hemi-prolate-spheroid-on-a-plane (HPS) model, discussed earlier. In the large-$\sigma_{\mathrm{a}}$ limit, the corresponding apex sharpness coefficient has the asymptotic form given by Eq.~\eqref{eq:hep_asym}, which is a particular case of the more general logarithmic structure
\begin{equation}
c_{\mathrm{a}}(\sigma_{\mathrm{a}})
\approx
\frac{1}
{g_1\ln(\sigma_{\mathrm{a}})-g_2},
\end{equation}
where $g_1$ and $g_2$ are constants.

The logarithmic dependence in Eq.~\eqref{eq:hep_asym} arises from the exact solution of Laplace's equation in prolate spheroidal coordinates, where the exact solution for electrostatic potential involves a first-order Legendre function of the second kind~\cite{LandauEHD, Kom91}. The appearance of a similarly slow variation in the present HCP numerical data suggests that a logarithmic functional structure may also be appropriate in this case, despite the absence of a known exact analytical solution.

A simple expression satisfying all the required conditions is
\begin{equation}
F(\sigma_{\mathrm{a}})
\equiv
\alpha
\ln\!\left(
\frac{1+\beta\sigma_{\mathrm{a}}}
{1+\beta}
\right),
\label{eq:F_def}
\end{equation}
where $\alpha$ and $\beta$ are positive constants. Substitution of Eq.~\eqref{eq:F_def} into Eq.~\eqref{eq:ca_interp}
yields
\begin{equation}
c_{\mathrm{a}}(\sigma_{\mathrm{a}})
=
c_{\mathrm{L}}
+
\frac{3-c_{\mathrm{L}}}
{1+
\alpha
\ln\!\left(
\frac{1+\beta\sigma_{\mathrm{a}}}
{1+\beta}
\right)},
\label{eq:ca_final}
\end{equation}
with the apex FEF given by eq. \eqref{eq:gamma_gen}.

The values of $\alpha$ and $\beta$ are then determined, as described in Appendix 3, by a complicated fitting procedure to the numerical data for $c_{\rm{a}}(\sigma_{\rm{a}})$. The value of $c_{\rm{L}}$ also emerges from this process.

The resulting formulation thus contains only two fitting parameters, $\alpha$ and $\beta$, along with the derived constant $c_{\rm{L}}$. The values of these parameters emerge as
\begin{equation}
\alpha = 1.099(1),
\qquad
\beta = 1.255(6),
\qquad
c_{\rm{L}} = 0.2573(3),
\label{alphabeta}
\end{equation}
where the numbers in parentheses indicate the uncertainties in the
final digits.

The resulting  expression \eqref{eq:ca_final}, when used in Eq. \eqref{eq:gamma_gen}, reproduces the numerical values of $\gamma_{\rm{a}}(\sigma_{\rm{a}})$ over the interval
$1 \leq \sigma_{\mathrm{a}} \leq 10^{3}$ with relative deviations below
approximately $0.15 \ \%$ (see Fig.\ref{fig:ca_sigma}).

The success of this compact formulation indicates that the ASR-dependence of the HCP apex field enhancement factor is essentially governed by two
simple physical requirements: the exact hemispherical limit and the
logarithmically slow reduction in geometric sensitivity as the emitter
becomes increasingly tall and/or slender.

\section{Fitting-formula error comparisons}

This section compares the behaviour of the new fitting formula above for the HCP-model apex-FEF $\gamma_{\rm{a}}$ with three previously published power-type fitting formulas for $\gamma_{\rm{a}}$. From least accurate to most accurate, the origins of the equations used are as follows.

\begin{figure*}[t]
\centering
\includegraphics[width=0.7\linewidth]{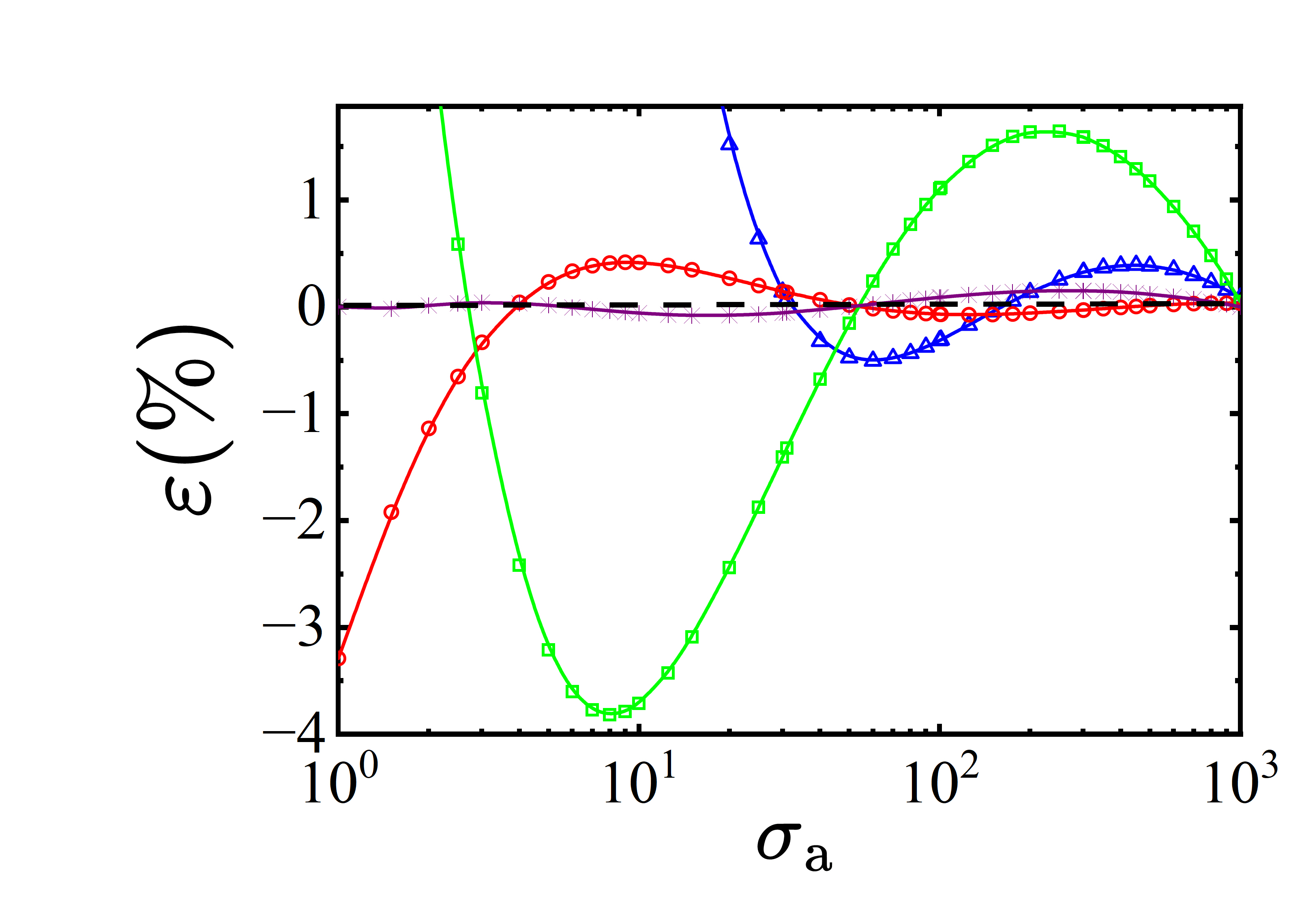}
\caption{Relative percentage deviation $\varepsilon(\sigma_{\mathrm a})$ between several fitting expressions and the high-precision finite-element reference results for the apex FEF shown in Fig.\ref{fig:ca_sigma}. The horizontal dashed line indicates zero relative deviation. Green squares correspond to the formula reported in Ref.~\cite{Edgcombe2001}; blue triangles to the expression proposed in Ref.~\cite{RBowring}; red circles to the approximation given in ~\cite{FEF2021}; and purple crosses to the present geometrically constrained formulation. The previously published power-type expressions exhibit systematic oscillatory deviations, reflecting their inability to reproduce the slowly varying large-$\sigma_{\mathrm a}$ behavior of $c_{\mathrm a}(\sigma_{\mathrm a})$. In contrast, the new expression maintains $|\varepsilon| \lesssim 0.15 \ \%$ (see Fig.\ref{fig:ca_sigma}) over the entire interval $1 \leq \sigma_{\mathrm a} \leq 10^{3}$, with no significant systematic excursions. This improvement results from explicitly incorporating the
exact hemispherical limit, monotonicity, and a logarithmic transition structure motivated by both the numerical HCP data and the exact hemi-prolate-spheroidal solution.}
\label{fig:error_comparison}
\end{figure*}

\begin{enumerate}
\item[(i)] 2001: Edgcombe+Valdre~\cite{Edgcombe2001} (green squares).
\item[(ii)] 2004: Read+Bowring~\cite{RBowring} eq.(2) (blue triangles).
\item[(iii)] 2021: Dall'Agnol et al.~\cite{FEF2021} eq.(13) (red circles).
\item[(iv)] 2026: New equations here (purple crosses).
\end{enumerate}
Edgcombe and Valdr\`{e} eq. (6) is also given as eq. (20) in Ref. \cite{Forbes2003}.

For cases (i)--(iii), the equations used all take the form of Eq.~\eqref{eq:old_fit}, but with parameters as listed in Table~\ref{tab:params_ranked}.
\begin{table}[ht]
\caption{Parameter values $(a,b,c,d)$ for previously published power-type fitting expressions, ordered from least accurate (i) to most accurate among them (iii).}
\label{tab:params_ranked}
\begin{ruledtabular}
\begin{tabular}{lcccc}
Case & $a$ & $b$ & $c$ & $d$ \\
\hline
(i) Edgcombe+Valdr\`{e}~\cite{Edgcombe2001} & 1.2 & 2.15 & 1 & 0.90 \\
(ii) Read+Bowring~\cite{RBowring} & 1.0782 & 4.7 & 1 & 0.9152 \\
(iii) Dall'Agnol et al.~\cite{FEF2021} & 0.86391 & 0.97756 & 0.52989 & 1.77667 \\
\end{tabular}
\end{ruledtabular}
\end{table}

Figure~\ref{fig:error_comparison} shows relative percentage deviations $\varepsilon(\sigma_{\rm{a}})$ (relative to the numerical data), defined via 
\begin{equation}
\varepsilon(\sigma_{\rm{a}})/ \%=100\times
\left(\frac{\gamma_{\rm{a}}^{(\mathrm{fit})}-\gamma_{\rm{a}}^{(\mathrm{num})}}
{\gamma_{\rm{a}}^{\rm{(num)}}}\right)
\end{equation}
lying within the interval $1 \le \sigma_a \le 1000$.

It is important to emphasize that the previously published formulas were not originally obtained under identical numerical or fitting conditions. They were derived using
different numerical methods, different ranges of apex sharpness ratio,
and different target accuracies. Consequently, the deviations shown in
Fig.~\ref{fig:error_comparison} should not be interpreted simply as a
ranking of the original numerical studies. Rather, they quantify how the
published fitting formulas perform when evaluated against the present
high-precision finite-element reference data derived for the interval
$1 \leq \sigma_{\mathrm a} \leq 10^{3}$.

The formula of Edgcombe and Valdr\`{e} ~\cite{Edgcombe2001} was based on
finite-element calculations and fitted over the approximate interval
$4 \leq \sigma_{\mathrm a} \leq 3000$. The formula of Read and
Bowring~\cite{RBowring} was obtained using a boundary-element approach
combined with fitting over the approximate interval
$31 \leq \sigma_{\mathrm a} \leq 3001$, with a reported numerical
accuracy of order $0.1\%$. The four-parameter expression of
Dall'Agnol \textit{et al.}~\cite{FEF2021} was based on finite-element
calculations using a specialized numerical methodology and was fitted
over the interval
$1 \leq \sigma_{\mathrm a} \leq 1000$.

In the earlier publications, the
number and distribution of fitting points were not always specified.
Moreover, their reported maximum deviations refer to their own numerical
datasets and fitting intervals, not to the present reference dataset.

It is also important to note that the use of power-type fitting
expressions is mathematically legitimate as a finite-interval
interpolation strategy. Although logarithmic and power-law dependences
possess fundamentally different asymptotic structures, a sufficiently
flexible power-law form can still reproduce a slowly varying function
with reasonable accuracy over a finite interval. In particular, the four-parameter expression proposed in
Ref.~\cite{FEF2021} contains enough adjustable freedom to mimic the
gradual evolution of
$c_{\mathrm a}(\sigma_{\mathrm a})$ over the interval investigated
there. Its relatively good performance is therefore not surprising,
especially since the effective exponent governing the large-
$\sigma_{\mathrm a}$ behaviour remains close to unity, leading to a
slow variation of the corresponding geometric coefficient.

However, when high precision is required simultaneously over several
decades in $\sigma_{\mathrm a}$, the structural differences between
power-law and logarithmic-type behaviour become increasingly visible in
the residual error patterns.

The earlier fitting formulae were not constructed to reproduce the
exact hemispherical limit at $\sigma_{\mathrm a}=1$, and some were
fitted only for comparatively large values of
$\sigma_{\mathrm a}$. Consequently, larger deviations near the lower end
of the present interval are not unexpected. In particular, the formula
of Edgcombe and Valdr\`{e}  exhibits its largest deviation near
$\sigma_{\mathrm a}=1$, where the original fitting procedure imposed no
hemispherical-limit constraint. The present expression, by contrast, was fitted over the full interval
$1 \leq \sigma_{\mathrm a} \leq 10^{3}$ using high-precision
finite-element data with relative numerical uncertainty in
$\gamma_{\mathrm a}$ of order $10^{-5}  \%$. In addition, the exact
hemispherical condition $c_{\mathrm a}(1)=3$ was explicitly enforced in
the analytical construction.

With these qualifications in mind,
Fig.~\ref{fig:error_comparison} shows clear differences in the residual
structure of the fitting formulas. 

These systematic differences are consistent with the limited asymptotic
flexibility of fixed-exponent power-law structures when attempting to
represent the gradual logarithmic-type evolution of
$c_{\mathrm a}(\sigma_{\mathrm a})$ observed in the present
high-precision numerical data over three decades in
$\sigma_{\mathrm a}$. The improved behaviour of the present expression
follows from incorporating directly into the analytical structure the
exact hemispherical limit, monotonicity, and the logarithmic transition
suggested both by the numerical HCP data and by the exact asymptotic
structure of the hemi-prolate-spheroidal solution.

\section{Conclusions}

The present work develops a geometrically constrained analytical
formulation for the apex sharpness coefficent and hence for the apex field enhancement factor of the hemisphere-on-cylindrical-post (HCP) model, in the large-gap regime.
High-precision axisymmetric finite-element calculations performed using
the Minimum Domain Dimension technique provided reference data with
relative numerical uncertainty of order $10^{-5} \ \%$ over the interval
$1 \leq \sigma_{\mathrm a} \leq 10^{3}$. The numerical results show that the apex field enhancement factor is primarily governed by the apex sharpness ratio
$\sigma_{\mathrm a}=h/r_{\mathrm a}$ and that the related
geometric coefficient $c_{\rm{a}}$ decreases monotonically from the exact
hemispherical limit while evolving only slowly for large
$\sigma_{\mathrm a}$. The numerical behaviour further indicates that the
large-$\sigma_{\mathrm a}$ regime is more naturally described by a
logarithmic-type dependence than by a fixed-exponent power law.

Motivated by these observations, and by the exact asymptotic structure
of the hemi-prolate-spheroidal solution, a logarithmic transition
function was incorporated into the analytical construction. The
resulting expression reproduces the finite-element reference data with
relative deviations below approximately $0.15 \ \%$ over the full interval
considered here. In addition, the analytical derivative remains in
close agreement with the numerical derivative extracted from the
finite-element data, providing a stringent consistency test of the
functional structure of the formulation. Comparison with previously published power-type fitting expressions
shows that such formulas remain useful as finite-interval interpolation
strategies, particularly when several adjustable parameters are
employed. However, their fixed-exponent asymptotic structure leads to
systematic residual patterns when compared against the present
high-precision numerical data over three decades in
$\sigma_{\mathrm a}$. By contrast, the present logarithmic formulation
produces substantially smoother and smaller residual deviations across
the full interval.

The present analysis therefore indicates that the HCP apex field enhancement factor is governed primarily by three structural features: the exact hemispherical limit, monotonic evolution of the geometric coefficient $c_{\rm{a}}$, and logarithmically slow variation in the large-$\sigma_{\mathrm a}$ regime. The formulation developed here provides a compact and quantitatively accurate analytical representation of the HCP apex-FEF. Further, it establishes a structurally consistent framework for future classical analytical and semi-analytical investigations of electrostatic field enhancement in emitter geometries for which exact analytical solutions are not currently known. We note, however, that all of the discussion in this paper is based on classical electrostatic theory. Eventually, this will need to be linked to a quantum-mechanical theory of charged surfaces, which will presumably result in some (relatively small) corrections. For example, some of the near-surface complications associated with using a quantum-mechanical theory of charged metal surfaces have been addressed by Lepetit \cite{Lepetit2017} and by Y.M. Li et al. \cite{Rosenz23}. It will also be useful for future work to explore possible links between this work and the ``universal formula for the field enhancement factor'' developed by Biswas \cite{BiswasFEF}. More generally, we believe that this work will be useful, not only in the various contexts in which field electron emission occurs, but also in various contexts \cite{RGFTerra23} in so-called ``terrestrial electrostatics", including in particular the theory of field enhancement by real-life lightning rods.

\section*{APPENDIX 1  \break \break  ELECTROSTATICS OF THE HEMISPHERE-ON-A-PLANE}

\begin{figure}
	\centering
	\includegraphics[width=0.8\linewidth]{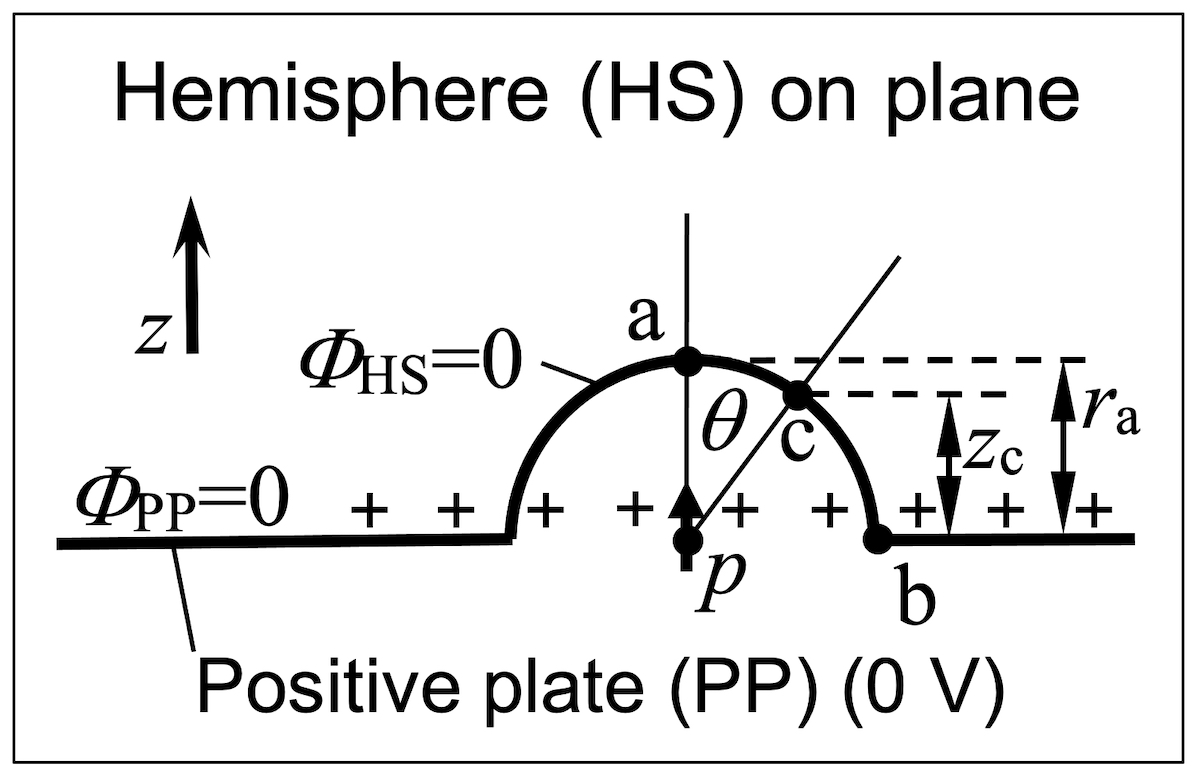}
    \caption{To illustrate the ``hemisphere-on-a-plane" field-enhancement model, in the case where the emitter is charged positively. For detailed explanation of the symbols used to label the diagram, see text.} 
\label{fig:HS}
\end{figure}

This Appendix considers the system, shown in Fig.~\ref{fig:HS}, that consists of a hemisphere on a plane of large lateral extent, for the situation where there would be an uniform ``background" electrostatic (ES) field $E_{\rm{back}}$ above the plane in the absence of the hemisphere. It will be shown that the ES field $E_{\rm{a}}$ at the apex of the hemisphere is equal to $3E_{\rm{back}}$, and hence that the apex field enhancement factor $\gamma_{\rm{a}}$ is equal to 3 precisely. This result also applies to the situation of a complete sphere in an uniform ES field.

In classical ES theory, this result applies to both positive and negative background fields. However, the situation of a positively charged planar plate is easier to discuss clearly, so this is done here.

Although there are long-established discussions in the literature based on general theoretical considerations, for example those of Jeans \cite{Jeans} (see his Section 319), there is also a simple physically direct proof, given below, that is difficult to find in textbooks.

The simplifying assumption is first made that all surfaces in the system have the same local work function, and hence that the ES potential $\mathit{\Phi}$  is the same at all points ``just outside" the surfaces of the positively charged plate and the hemisphere.

The hemisphere is then modeled by placing, at the centre of its base, a (positive) electric dipole of strength $p$. (That is, the positive end of the dipole is on the positive-$z$ side of the plate.) 

The ES potential $\mathit{\Phi_{\rm{d}}}$ due to this dipole has the well known form
\begin{equation}
\mathit{\Phi_{\rm{d}}}(r,\theta)  =  p \mathrm{cos} \theta /4 \pi \epsilon_{\mathrm{0}} r^2,
\label{eq:phidef}
\end{equation}
where $r$ is distance from the dipole centre and $\theta$ is the angle shown in Fig.~\ref{fig:HS}. It follows that the resulting electrotatic potential difference (ESPD) $\Delta \mathit{\Phi} [\rm{dipole}]$ between points
``a" and ``b" is
\begin{equation}
\Delta \mathit{\Phi} [\rm{dipole}]  =  (\mathit{p}/4 \pi \epsilon_{\rm{0}} \mathit{r}_{\rm{a}}^2) [\mathrm{cos0} - \mathrm{cos}(\pi/2)]  =  (\mathit{p}/4 \pi \epsilon_{\rm{0}} \mathit{r}_{\rm{a}}^2).   
\label{eq:dipole}
\end{equation}

The ESPD $\Delta \mathit{\Phi} [\rm{back}]$ between points
``a" and ``b" that results from the (positive) background field is
\begin{equation}
\Delta \mathit{\Phi} [\rm{back}]  = - \mathit{E}_{\rm{back}} \mathit{r}_{\rm{a}}.  
\label{eq:backfield}
\end{equation}
Hence, the total ESPD  $\Delta \mathit{\Phi} [\rm{total}]$ between points
``a" and ``b" is
\begin{equation}
\Delta \mathit{\Phi} [\rm{total}]  = (\mathit{p}/4 \pi \epsilon_{\rm{0}} \mathit{r}_{\rm{a}}^2) - \mathit{E}_{\rm{back}} \mathit{r}_{\rm{a}}.  
\label{eq:total}
\end{equation}
This ESPD must be equal to zero, which implies that
\begin{equation}
\mathit{p}/4 \pi \epsilon_{\rm{0}} \mathit{r}_{\rm{a}}^3 = \mathit{E}_{\rm{back}}.  
\label{eq:field1}
\end{equation}

The (positive) ES field at ``a" due to the dipole is $2 p /4 \pi \epsilon_{\rm{0}} \mathit{r}_{\rm{a}}^3$, which equals $2 \mathit{E}_{\rm{back}}$. The background field adds to this, giving the total field $E_{\rm{a}}$ at ``a" as $3 \mathit{E}_{\rm{back}}$.

It follows that the apex FEF $\gamma_{\rm{a}} [\equiv E_{\rm{a}} / E_{\rm{back}} ]$ is 3 precisely. Since the apex sharpness ratio $\sigma_{\rm{a}}$ for a hemisphere equals 1, it further follows that, for a hemisphere on a plane, the apex sharpness coefficient $c_{\rm{a}} [\equiv \gamma_{\rm{a}} / \sigma_{\rm{a}} ]$ is also equal to 3 precisely. 

For consistency, it remains to show that this dipole-based model ensures that the whole of the hemisphere surface is at zero ES potential. From Eq.~\eqref{eq:phidef}, the ES potential at ``c", relative to ``b", is $p \mathrm{cos} \theta /4 \pi \epsilon_{\mathrm{0}} r_{\rm{a}}^2$, which from Eq.~\eqref{eq:total} is equal to
$\mathit{E}_{\rm{back}} \mathit{r}_{\rm{a}} \mathrm{cos} \theta $. But the potential contribution from the background field is $\mathit{E}_{\rm{back}} z_{\rm{c}} = -E_{\rm{back}} \mathit{r}_{\rm{a}} \mathrm{cos} \theta$. Thus, the two contributions cancel and the ES potential at ``c" is zero. This argument applies to any point on the hemisphere surface.

The importance of this ``HS" result is that it forms a limiting result for $\gamma_{\rm{a}}$ and $c_{\rm{a}}$  to which all formulae for cylindricallly symmetric posts with a hemispherically rounded apex should reduce for $\sigma_{\rm{a}}=1$.

\section*{APPENDIX 2 \break \break APEX SHARPNESS RATIO}

This Appendix shows that the apex sharpness ratio (ASR) $\sigma_{\rm{a}}$ is physically the most significant parameter (rather than, say, the aspect ratio), when discussing electrostatic field enhancement generated by cylindically symmetric post-like structures with a smooth rounded apex.

\begin{figure}
	\centering
	\includegraphics[width=0.6\linewidth]{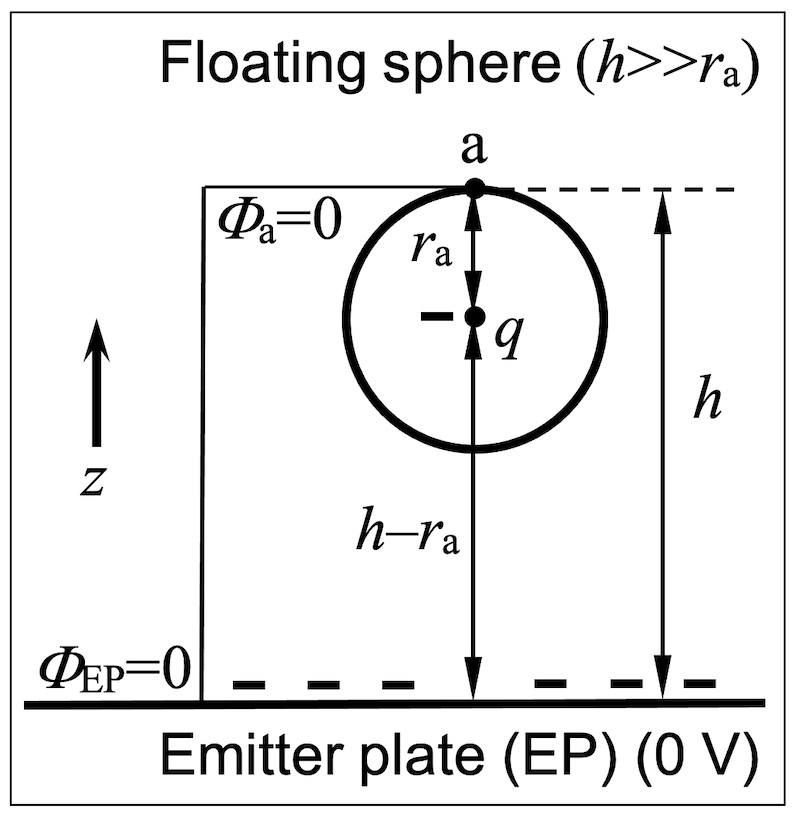}
    \caption{To illustrate the ``floating sphere at emitter plate potential" field-enhancement model, in the case where the emitter is charged negatively. The diagram is not to scale, particularly vertically. For explanation of the symbols used to label the diagram, see text.} 
\label{fig:Float}
\end{figure}

The simplest model for such a structure has long been known \cite{Vibrans1964} to be the ``floating sphere at emitter plate potential", as illustrated in Fig. \ref{fig:Float}.  The analysis here is a slightly modified version of that given in an earlier paper \cite{RFJAP2016}, and applies to the situation where the emitter is charged \textit{negatively}.

In the electrical system shown, the requirement for the electrons to be in thermodynamic/statistical-mechanical equilibrium is that the electron Fermi level at the apex ``a" of the floating sphere be equal to the electron Fermi level in the planar emitter plate (EP). Or, in classical electrical-engineering terms, that the measured voltage between ``a" and ``EP" be zero.

The simplifying assumption is then made that all surfaces in the system have the same local work function. The above requirement then reduces to the requirement that the electrostatic potential (ES) ${\mathit{\Phi}}_{\rm{a}}$ just outside the sphere apex be equal to the ES potential ${\mathit{\Phi}}_{\rm{EP}}$ just outside the emitter plate.

Due to the (negative) background ES field $E_{\rm{back}}$, there would be a positive ES potential difference $\Delta \mathit{\Phi} [\rm{field]}$ between ``a" and ``EP" given by 
$\Delta \mathit{\Phi} [\rm{field}] = -\mathit{E}_{\rm{back}}\mathit{h}$.

This can be compensated by replacing the sphere by a (negative) point charge $q$ at the sphere centre. This creates a (negative) ES potential difference $\Delta \mathit{\Phi} [\rm{charge]}$ between ``a" and ``EP" given by
\begin{equation}
\Delta \mathit{\Phi} [\rm{charge]} = \mathit{q}/4 \pi \epsilon_0 \mathit{r}_{\rm{a}} 
- \mathit{q}/4 \pi \epsilon_0 (\mathit{h} - \mathit{r}_{\rm{a}}).
\label{eq:deltaq}
\end{equation}

Since $h \gg r_{\rm{a}}$, the second term may be neglected, and the combined effect of the ``field" and ``charge" terms can be written as
\begin{equation}
\Delta \mathit{\Phi} [\rm{combined]} =  - \mathit{E}_{\rm{back}}\mathit{h}  +  \mathit{q}/4 \pi \epsilon_0 \mathit{r}_{\rm{a}} = 0 ,
\label{eq:deltaboth}
\end{equation}
which shows that
\begin{equation}
\mathit{q}/4 \pi \epsilon_0 \mathit{r}_{\rm{a}} =  \mathit{E}_{\rm{back}}\mathit{h}.
\label{eq:deltaboth2}
\end{equation}

At point ``a", the (negative) field $E_{\rm{a}}[\rm{charge}]$ due to (negative) charge $q$ is given by
\begin{equation}
E_{\rm{a}}[\rm{charge}] = \mathit{q}/4 \pi \epsilon_0 {\mathit{r}_{\rm{a}}}^2
=  \mathit{E}_{\rm{back}} (\mathit{h} / \mathit{r}_{\rm{a}}).
\label{eq:deltaboth3}
\end{equation}
At ``a" there is also a field contribution due to the background field, but this is relatively small when $h/\mathit{r}_{\rm{a}} \gg 1$, and can be neglected.

Hence, in this simple model, the formula for the apex field enhancement factor (apex FEF) $\gamma_{\rm{a}}$ becomes 
\begin{equation}
\gamma_{\rm{a}}  =  E_{\rm{a}} / E_{\rm{back}} = \mathit{h} / \mathit{r}_{\rm{a}}
=  \sigma_{\rm{a}}.
\label{eq:gamma}
\end{equation}
An equivalent treatment using a positively charged emitter confirms that, in the classical theory used here, eq. \eqref{eq:gamma} applies to both emitter polarities.

The purpose of this treatment has been to confirm what the ``leading factor" is in the expression for apex FEF. It has been shown that this leading factor is the \textit{apex sharpness ratio} $\sigma_{\rm{a}} \  [=h/r_{\rm{a}}]$. Clearly, many small terms have been omitted, and clearly the floating sphere is not a detailed model for posts and protrusions. Thus, it is to be expected that, in detailed expressions for apex FEFs, a correction factor (our $c_{\rm{a}}$) would also be needed.

It might also reasonably be expected that the value of $c_{\rm{a}}$  would change more rapidly as $h$ diminishes towards $r_{\rm{a}}$.

\section*{APPENDIX 3 \break \break DETAILED MATHEMATICAl ARGUMENTS}

\subsection{Principles}

The function defined by eq. \eqref{eq:F_def} was
\begin{equation}
F(\sigma_{\mathrm{a}})
\equiv
\alpha
\ln\!\left(
\frac{1+\beta\sigma_{\mathrm{a}}}
{1+\beta}
\right),
\label{eq:F_def2}
\end{equation}
where $\alpha$ and $\beta$ are positive constants.

This function satisfies the following conditions:

\begin{enumerate}
\item $F(1)=0$ exactly;
\item $F(\sigma_{\mathrm{a}})$ is strictly increasing for
      $\sigma_{\mathrm{a}}>1$;
\item For large $\sigma_{\mathrm{a}}$,
\begin{equation}
F(\sigma_{\mathrm{a}}) \approx \alpha\ln(\sigma_{\mathrm{a}}) + K,
\end{equation}
where $K$is an initially unknown constant given by
\begin{equation}
K = \alpha
\ln\!\left(
\frac{\beta}{1+\beta}
\right).
\end{equation}
\end{enumerate}
%

Substituting the above asymptotic form into Eq.~\eqref{eq:ca_interp} yields a large-$\sigma_{\rm{a}}$ form for $c_{\rm{a}}(\sigma_{\rm{a}})$, for the HCP model, namely  
\begin{equation}
c_{\mathrm{a}}(\sigma_{\mathrm{a}}) \approx c_{\mathrm{L}} + \frac{3-c_{\mathrm{L}}}
{\alpha\ln(\sigma_{\mathrm{a}})+K+1}.
\label{eq:ca_asym}
\end{equation}

This has the general form
\begin{equation}
c_{\mathrm{a}}(\sigma_{\mathrm{a}}) - c_{\mathrm{L}} \approx
\frac{1} {g_1\ln(\sigma_{\mathrm{a}})-g_2},
\label{eq:ca_residual}
\end{equation}
with
\begin{equation}
g_1
\equiv
\frac{\alpha}{3-c_{\mathrm{L}}},
\qquad
g_2
\equiv
-\frac{K+1}{3-c_{\mathrm{L}}}.
\end{equation}

Equation~\eqref{eq:ca_residual} shows that, in the HCP model, as derived from numerical simulations over the range $1 \leq \sigma_{\rm{a}} \leq 1000$, $c_{\rm{a}}(\sigma_{\rm{a}})$ reduces towards $c_{\rm{L}}$, as $\sigma_{\rm{a}}$ becomes large, in a logarithmic way that is formally similar to the way that, in the HPS model, $c_{\rm{a}}(\sigma_{\rm{a}})$ reduces towards 0, as $\sigma_{\rm{a}}$ becomes large.

In the HPS model, $\sigma_{\rm{a}}$ tends to infinity linearly, whereas $c_{\rm{a}}(\sigma_{\rm{a}})$ tends to zero logaritmically; hence the apex FEF $\gamma_{\rm{a}}(\sigma_{\rm{a}}) = $$c_{\rm{a}}(\sigma_{\rm{a}}) \cdot \sigma_{\rm{a}}$ tends to infinity as $\sigma_{\rm{a}}$ tends to infinity.

At present, for the HCP model, the limitations imposed by our finite simulation range mean that we do not know whether the value of $c_{\rm{a}}$ at infinity would be zero or finite. Nevertheless, what can be concluded is that the approach to the weakly varying large-$\sigma_{\rm{a}}$ regime is governed by a logarithmic-type functional structure analogous to that arising analytically in the exact HPS-model solution.

This correspondence is significant because it shows that the logarithmic
term in Eq.~\eqref{eq:F_def} is not introduced solely as an empirical
fitting device. Rather, it reflects a mathematical structure that is
already known to emerge in an exactly solvable electrostatic problem
closely related to the HCP geometry. In this sense, the present
expression may be viewed as an obvious variation of the exact HPS
functional form to suit a geometry for which no closed analytical solution is
currently available.

\subsection{Determination of values for constants}

It remains to explain how the values of $\alpha$ and $\beta$, and related parameters, are determined.

Inverting Eq. \eqref{eq:ca_interp2}, to give an expression for $c_{\rm{L}}$ in terms of $c_{\rm{a}}(1000)$, yields
\begin{equation}
c_{\mathrm{L}}
=
c_{\mathrm{a}}(1000)
+
\frac{c_{\mathrm{a}}(1000)-3}
{F(1000)}.
\label{eq:cLnum}
\end{equation}
Setting $\sigma_{\mathrm{a}}=1000$ in Eq.~\eqref{eq:F_def2}, and then inserting the resulting expression for $F(1000)$ into Eq.~\eqref{eq:cLnum}, yields
\begin{equation}
c_{\mathrm{L}}
=
c_{\mathrm{a}}(1000)
+
\frac{c_{\mathrm{a}}(1000)-3}
{\alpha
\ln\!\left(
\frac{1+1000\beta}
{1+\beta}
\right)}.
\label{eq:cL_explicit}
\end{equation}
\\
Thus, $c_{\mathrm{L}}$ is not an additional fitting parameter:
Eq.~\eqref{eq:cL_explicit} expresses $c_{\mathrm{L}}$ explicitly in terms of the two
parameters $\alpha$ and $\beta$ that define the transition function.

Substitution of Eq.~\eqref{eq:cL_explicit} into Eq.~\eqref{eq:ca_interp} yields a rather complicated expression (not given here) that contains $\alpha$ and $\beta$ as the only adjustable parameters  needed.  Their values are then determined by least-squares fit to the finite-element data, as indicated in the main text, and $c_{\mathrm{L}}$ is obtained from eq. \eqref{eq:cL_explicit}.

The resulting values of $\alpha$, $\beta$ and $c_{\mathrm{L}}$ are stated as Eq.(\ref{alphabeta})]. The related values of the other constants appearing in this Appendix are as follow
\begin{equation}
K=-0.6439(2),
\end{equation}
and
\begin{equation}
g_1=0.4008(5),
\qquad
g_2=-0.1298(7).
\end{equation}
As earlier, the figures in brackets indicate the uncertainty in the least significant figure.

\section*{Acknowledgments}

TAdA thanks the Conselho Nacional de Desenvolvimento Científico e Tecnológico (CNPq), Grant No. 305688/2023-5.

\section*{Data Availability}

The datasets supporting this article are available from the Zenodo repository at~\cite{dataset2026}.

\bibliography{Refs_MF_WV.bib}

\end{document}